# Embedded Software of the KM3NeT Central Logic Board

S. Aiello[a], A. Albert[b,bd], S. Alves Garre[c], Z. Aly[d], A. Ambrosone[f,e], F. Ameli[g], M. Andre[h], E. Androutsou[i], M. Anghinolfi[j], M. Anguita[k], L. Aphecetche[l], M. Ardid[m], S. Ardid[m], H. Atmani[n], J. Aublin[o], C. Bagatelas[i], L. Bailly-Salins[p], Z. Bardačová[r,q], B. Baret[o], S. Basegmez du Pree[s], Y. Becherini[o], M. Bendahman[n,o], F. Benfenati[u,t], M. Benhassi[v,e], D. M. Benoit[w], E. Berbee[s], V. Bertin[d], V. van Beveren[s,*], S. Biagi[x], M. Boettcher[y], J. Boumaaza[n], M. Bouta[z], M. Bouwhuis[s], C. Bozza[aa,e], R. M. Bozza[f,e], H.Brânzaş[ab], F. Bretaudeau[l], R. Bruijn[ac,s], J. Brunner[d], R. Bruno[a], E. Buis[ad,s], R. Buompane[v,e], J. Busto[d], B. Caiffi[j], D. Calvo[c], S. Campion[g,ae], A. Capone[g,ae], F. Carenini[u,t], V. Carretero[c], T. Cartraud[o], P. Castaldi[af,t], V. Cecchini[c], S. Celli[g,ae], L. Cerisy[d], M. Chabab[ag], M. Chadolias[ah], A. Chen[ai], S. Cherubini[aj,x], T. Chiarusi[t], M. Circella[ak], R. Cocimano[x], J. A. B. Coelho[o], A. Coleiro[o], R. Coniglione[x], P. Coyle[d], A. Creusot[o], A. Cruz[al], G. Cuttone[x], R. Dallier[l], Y. Darras[ah], A. De Benedittis[e], B. De Martino[d], V. Decoene[l], R. Del Burgo[e], L. S. Di Mauro[x], I. Di Palma[g,ae], A. F. Díaz[k], D. Diego-Tortosa[x], C. Distefano[x], A. Domi[ac,s], C. Donzaud[o], D. Dornic[d], M. Dörr[am], E. Drakopoulou[i], D. Drouhin[b,bd], R. Dvornický[r], T. Eberl[ah], E. Eckerová[r,q], A. Eddymaoui[n], T. van Eeden[s], M. Eff[ah], D. van Eijk[s], I. El Bojaddaini[z], S. El Hedri[o], A. Enzenhöfer[d], G. Ferrara[x], M. D. Filipović[an], F. Filippini[u,t], L. A. Fusco[aa], O. Gabella[ao], J. Gabriel[ap], S. Gagliardini[g], T. Gal[ah], J. García Méndez[m], A. Garcia Soto[c], C. Gatius Oliver[s], N. Geißelbrecht[ah], H. Ghaddari[z], L. Gialanella[e,v], B. K. Gibson[w], E. Giorgio[x], A. Girardi[g], I. Goos[o], D. Goupilliere[p], S. R. Gozzini[c], R. Gracia[ah], K. Graf[ah], C. Guidi[aq,j], B. Guillon[p], M. Gutiérrez[ar], H. van Haren[as], A. Heijboer[s], A. Hekalo[am], L. Hennig[ah], J. J. Hernández-Rey[c], F. Huang[d], W. Idrissi Ibnsalih[e], G. Illuminati[u,t], C. W. James[al], P. Jansweijer[s], M. de Jong[at,s], P. de Jong[ac,s], B. J. Jung[s], P. Kalaczyński[au,be], O. Kalekin[ah], U. F. Katz[ah], N. R. Khan Chowdhury[c], A. Khatun[r], G. Kistauri[aw,av], C. Kopper[ah], A. Kouchner[ax,o], V. Kulikovskiy[j], R. Kvatadze[aw],

---

*corresponding author

*Email addresses:* km3net-pc@km3net.de; v.van.beveren@nikhef.nl (V. van Beveren), real@ific.uv.es (D. Real)

M. Labalme[p], R. Lahmann[ah], G. Larosa[x], C. Lastoria[d], A. Lazo[c], S. Le Stum[d], G. Lehaut[p], E. Leonora[a], N. Lessing[c], G. Levi[u,t], M. Lindsey Clark[o], F. Longhitano[a], J. Majumdar[s], L. Malerba[j], F. Mamedov[q], J. Mańczak[c], A. Manfreda[e], M. Marconi[aq,j], A. Margiotta[u,t], A. Marinelli[e,f], C. Markou[i], L. Martin[l], J. A. Martínez-Mora[m], F. Marzaioli[v,e], M. Mastrodicasa[ae,g], S. Mastroianni[e], S. Miccichè[x], G. Miele[f,e], P. Migliozzi[e], E. Migneco[x], S. Minutoli[j], M. L. Mitsou[e], C. M. Mollo[e], L. Morales-Gallegos[v,e], C. Morley-Wong[al], A. Mosbrugger[ah], A. Moussa[z], I. Mozun Mateo[az,ay], R. Muller[s], M. R. Musone[e,v], M. Musumeci[x], L. Nauta[s], S. Navas[ar], A. Nayerhoda[ak], C. A. Nicolau[g], B. Nkosi[ai], B. Ó Fearraigh[ac,s], V. Oliviero[f,e], A. Orlando[x], E. Oukacha[o], J. Palacios González[c], G. Papalashvili[av], E.J. Pastor Gomez[c], A. M. Păun[ab], G. E. Păvălaş[ab], S. Peña Martínez[o], M. Perrin-Terrin[d], J. Perronnel[p], V. Pestel[az], R. Pestes[o], P. Piattelli[x], C. Poirè[aa], V. Popa[ab], T. Pradier[b], S. Pulvirenti[x], G. Quéméner[p], C. Quiroz[m], U. Rahaman[c], N. Randazzo[a], S. Razzaque[ba], I. C. Rea[e], D. Real[c,*], S. Reck[ah], G. Riccobene[x], J. Robinson[y], A. Romanov[aq,j], A. Saina[c], F. Salesa Greus[c], D. F. E. Samtleben[at,s], A. Sánchez Losa[c,ak], M. Sanguineti[aq,j], C. Santonastaso[bb,e], D. Santonocito[x], P. Sapienza[x], Y. Scarpetta[aa], J. Schnabel[ah], M. F. Schneider[ah], J. Schumann[ah], H. M. Schutte[y], J. Seneca[s], B. Setter[ah], I. Sgura[ak], R. Shanidze[av], Y. Shitov[q], F. Šimkovic[r], A. Simonelli[e], A. Sinopoulou[a], M.V. Smirnov[ah], B. Spisso[e], M. Spurio[u,t], D. Stavropoulos[i], I. Štekl[q], M. Taiuti[aq,j], Y. Tayalati[n], H. Tedjditi[j], H. Thiersen[y], I. Tosta e Melo[a,aj], B. Trocme[o], S. Tsagkli[i], V. Tsourapis[i], E. Tzamariudaki[i], A. Vacheret[p], V. Valsecchi[x], V. Van Elewyck[ax,o], G. Vannoye[d], G. Vasileiadis[ao], F. Vazquez de Sola[s], C. Verilhac[o], A. Veutro[g,ae], S. Viola[x], D. Vivolo[v,e], H. Warnhofer[ah], J. Wilms[bc], E. de Wolf[ac,s], H. Yepes-Ramirez[m], G. Zarpapis[i], S. Zavatarelli[j], A. Zegarelli[g,ae], D. Zito[x], J. D. Zornoza[c], J. Zúñiga[c], N. Zywucka[y]

[a]*INFN, Sezione di Catania, Via Santa Sofia 64, Catania, 95123 Italy*

[b]*Université de Strasbourg, CNRS, IPHC UMR 7178, F-67000 Strasbourg, France*

[c]*IFIC - Instituto de Física Corpuscular (CSIC - Universitat de València), c/Catedrático José Beltrán, 2, 46980 Paterna, Valencia, Spain*

[d]*Aix Marseille Univ, CNRS/IN2P3, CPPM, Marseille, France*

[e]*INFN, Sezione di Napoli, Complesso Universitario di Monte S. Angelo, Via Cintia ed. G, Napoli, 80126 Italy*

[f]*Università di Napoli "Federico II", Dip. Scienze Fisiche "E. Pancini", Complesso Universitario di Monte S. Angelo, Via Cintia ed. G, Napoli, 80126 Italy*

[g]*INFN, Sezione di Roma, Piazzale Aldo Moro 2, Roma, 00185 Italy*

[h] *Universitat Politècnica de Catalunya, Laboratori d'Aplicacions Bioacústiques, Centre Tecnològic de Vilanova i la Geltrú, Avda. Rambla Exposició, s/n, Vilanova i la Geltrú, 08800 Spain*
[i] *NCSR Demokritos, Institute of Nuclear and Particle Physics, Ag. Paraskevi Attikis, Athens, 15310 Greece*
[j] *INFN, Sezione di Genova, Via Dodecaneso 33, Genova, 16146 Italy*
[k] *University of Granada, Dept. of Computer Architecture and Technology/CITIC, 18071 Granada, Spain*
[l] *Subatech, IMT Atlantique, IN2P3-CNRS, Université de Nantes, 4 rue Alfred Kastler - La Chantrerie, Nantes, BP 20722 44307 France*
[m] *Universitat Politècnica de València, Instituto de Investigación para la Gestión Integrada de las Zonas Costeras, C/ Paranimf, 1, Gandia, 46730 Spain*
[n] *University Mohammed V in Rabat, Faculty of Sciences, 4 av. Ibn Battouta, B.P. 1014, R.P. 10000 Rabat, Morocco*
[o] *Université Paris Cité, CNRS, Astroparticule et Cosmologie, F-75013 Paris, France*
[p] *LPC CAEN, Normandie Univ, ENSICAEN, UNICAEN, CNRS/IN2P3, 6 boulevard Maréchal Juin, Caen, 14050 France*
[q] *Czech Technical University in Prague, Institute of Experimental and Applied Physics, Husova 240/5, Prague, 110 00 Czech Republic*
[r] *Comenius University in Bratislava, Department of Nuclear Physics and Biophysics, Mlynska dolina F1, Bratislava, 842 48 Slovak Republic*
[s] *Nikhef, National Institute for Subatomic Physics, PO Box 41882, Amsterdam, 1009 DB Netherlands*
[t] *INFN, Sezione di Bologna, v.le C. Berti-Pichat, 6/2, Bologna, 40127 Italy*
[u] *Università di Bologna, Dipartimento di Fisica e Astronomia, v.le C. Berti-Pichat, 6/2, Bologna, 40127 Italy*
[v] *Università degli Studi della Campania "Luigi Vanvitelli", Dipartimento di Matematica e Fisica, viale Lincoln 5, Caserta, 81100 Italy*
[w] *E. A. Milne Centre for Astrophysics, University of Hull, Hull, HU6 7RX, United Kingdom*
[x] *INFN, Laboratori Nazionali del Sud, Via S. Sofia 62, Catania, 95123 Italy*
[y] *North-West University, Centre for Space Research, Private Bag X6001, Potchefstroom, 2520 South Africa*
[z] *University Mohammed I, Faculty of Sciences, BV Mohammed VI, B.P. 717, R.P. 60000 Oujda, Morocco*
[aa] *Università di Salerno e INFN Gruppo Collegato di Salerno, Dipartimento di Fisica, Via Giovanni Paolo II 132, Fisciano, 84084 Italy*
[ab] *ISS, Atomistilor 409, Măgurele, RO-077125 Romania*
[ac] *University of Amsterdam, Institute of Physics/IHEF, PO Box 94216, Amsterdam, 1090 GE Netherlands*
[ad] *TNO, Technical Sciences, PO Box 155, Delft, 2600 AD Netherlands*
[ae] *Università La Sapienza, Dipartimento di Fisica, Piazzale Aldo Moro 2, Roma, 00185 Italy*
[af] *Università di Bologna, Dipartimento di Ingegneria dell'Energia Elettrica e dell'Informazione "Guglielmo Marconi", Via dell'Università 50, Cesena, 47521 Italia*

[ag] *Cadi Ayyad University, Physics Department, Faculty of Science Semlalia, Av. My Abdellah, P.O.B. 2390, Marrakech, 40000 Morocco*
[ah] *Friedrich-Alexander-Universität Erlangen-Nürnberg (FAU), Erlangen Centre for Astroparticle Physics, Nikolaus-Fiebiger-Straße 2, 91058 Erlangen, Germany*
[ai] *University of the Witwatersrand, School of Physics, Private Bag 3, Johannesburg, Wits 2050 South Africa*
[aj] *Università di Catania, Dipartimento di Fisica e Astronomia "Ettore Majorana", Via Santa Sofia 64, Catania, 95123 Italy*
[ak] *INFN, Sezione di Bari, via Orabona, 4, Bari, 70125 Italy*
[al] *International Centre for Radio Astronomy Research, Curtin University, Bentley, WA 6102, Australia*
[am] *University Würzburg, Emil-Fischer-Straße 31, Würzburg, 97074 Germany*
[an] *Western Sydney University, School of Computing, Engineering and Mathematics, Locked Bag 1797, Penrith, NSW 2751 Australia*
[ao] *Laboratoire Univers et Particules de Montpellier, Place Eugène Bataillon - CC 72, Montpellier Cédex 05, 34095 France*
[ap] *IN2P3, LPC, Campus des Cézeaux 24, avenue des Landais BP 80026, Aubière Cedex, 63171 France*
[aq] *Università di Genova, Via Dodecaneso 33, Genova, 16146 Italy*
[ar] *University of Granada, Dpto. de Física Teórica y del Cosmos & C.A.F.P.E., 18071 Granada, Spain*
[as] *NIOZ (Royal Netherlands Institute for Sea Research), PO Box 59, Den Burg, Texel, 1790 AB, the Netherlands*
[at] *Leiden University, Leiden Institute of Physics, PO Box 9504, Leiden, 2300 RA Netherlands*
[au] *National Centre for Nuclear Research, 02-093 Warsaw, Poland*
[av] *Tbilisi State University, Department of Physics, 3, Chavchavadze Ave., Tbilisi, 0179 Georgia*
[aw] *The University of Georgia, Institute of Physics, Kostava str. 77, Tbilisi, 0171 Georgia*
[ax] *Institut Universitaire de France, 1 rue Descartes, Paris, 75005 France*
[ay] *IN2P3, 3, Rue Michel-Ange, Paris 16, 75794 France*
[az] *LPC, Campus des Cézeaux 24, avenue des Landais BP 80026, Aubière Cedex, 63171 France*
[ba] *University of Johannesburg, Department Physics, PO Box 524, Auckland Park, 2006 South Africa*
[bb] *Università degli Studi della Campania "Luigi Vanvitelli", CAPACITY, Laboratorio CIRCE - Dip. Di Matematica e Fisica - Viale Carlo III di Borbone 153, San Nicola La Strada, 81020 Italy*
[bc] *Friedrich-Alexander-Universität Erlangen-Nürnberg (FAU), Remeis Sternwarte, Sternwartstraße 7, 96049 Bamberg, Germany*
[bd] *Université de Haute Alsace, rue des Frères Lumière, 68093 Mulhouse Cedex, France*
[be] *AstroCeNT, Nicolaus Copernicus Astronomical Center, Polish Academy of Sciences, Rektorska 4, Warsaw, 00-614 Poland*

**Abstract**

The KM3NeT Collaboration is building and operating two deep sea neutrino telescopes at the bottom of the Mediterranean Sea. The telescopes consist of latices of photomultiplier tubes housed in pressure-resistant glass spheres, called digital optical modules and arranged in vertical detection units. The two main scientific goals are the determination of the neutrino mass ordering and the discovery and observation of high-energy neutrino sources in the Universe. Neutrinos are detected via the Cherenkov light, which is induced by charged particles originated in neutrino interactions. The photomultiplier tubes convert the Cherenkov light into electrical signals that are acquired and timestamped by the acquisition electronics. Each optical module houses the acquisition electronics for collecting and timestamping the photomultiplier signals with one nanosecond accuracy. Once finished, the two telescopes will have installed more than six thousand optical acquisition nodes, completing one of the more complex networks in the world in terms of operation and synchronization. The embedded software running in the acquisition nodes has been designed to provide a framework that will operate with different hardware versions and functionalities. The hardware will not be accessible once in operation, which complicates the embedded software architecture. The embedded software provides a set of tools to facilitate remote manageability of the deployed hardware, including safe reconfiguration of the firmware. This paper presents the architecture and the techniques, methods and implementation of the embedded software running in the acquisition nodes of the KM3NeT neutrino telescopes.

**Program Summary**

*Program title:* Embedded software for the KM3NeT CLB

*CPC Library link to program files:* (to be added by Technical Editor)

*Developer's repository link:* Restricted to KM3NeT users

*Code Ocean capsule:* (to be added by Technical Editor)

*Licensing provisions:* GNU General Public License 3 (GPL)

*Programming language:* C

*Nature of problem:* The challenge for the embedded software in the KM3NeT neutrino telescope lies in orchestrating the Digital Optical Modules (DOMs) to achieve the synchronized data acquisition of the incoming optical signals. The DOMs are the crucial component responsible for capturing neutrino interactions deep underwater. The embedded software must config-ure and precisely time the operation of each DOM. Any deviation or timing mismatch could compromise data integrity, undermining the scientific value of the experiment. Therefore, the embedded software plays a critical role in coordinating, synchronizing, and operating these modules, ensuring they work in unison to capture and process neutrino signals accurately, ultimately advancing our understanding of fundamental particles in the Universe.

*Solution method:* The embedded software on the DOMs provides a solution based on a C-based bare-metal application, operating without a real-time embedded OS. It is loaded into the RAM during FPGA configuration, consuming less than 256 kB of RAM. The software architecture comprises two layers: system software and application. The former offers OS-like features, including a multitasking scheduler, firmware updates, peripheral drivers, a UDP-based network stack, and error handling utilities. The application layer contains a state machine ensuring consistent program states. It is navigated via slow control events, including external inputs and autonomous responses. Subsystems within the application code control specific acquisition electronics components via the associated driver abstractions.

*Additional comments including Restrictions and Unusual features:* Due to the operation conditions of the neutrino telescope, where access is restricted, the embedded software implements a fail-safe procedure to reconfigure the the firmware were the embedded software runs.

*References:*

---

## 1. Introduction

The KM3NeT Collaboration is currently installing a research infrastructure at the bottom of the Mediterranean Sea [1]. The infrastructure comprises two neutrino detectors: Astroparticle Research with Cosmics in the Abyss (ARCA) [2] and Oscillation Research with Cosmics in the Abyss (ORCA) [3]. ARCA has been designed for the detection of neutrinos of astrophysical origin with energies from ∼100 GeV to PeV scale, and is located 100 km off the southern tip of Sicily, Italy, at a depth of about 3500 m. ORCA, which is optimized for studying fundamental properties of neutrinos, is located about 40 km south of the coast of Toulon, France, at a depth of about 2450 m. The neutrino detectors instrument very large de-tection volumes of seawater with three-dimensional arrays of light detectors, the Digital Optical Modules (DOMs) [4, 5], to detect Cherenkov light, which is induced in the seawater by charged particles generated in neutrino interactions (see Fig. 1). The DOM, a pressure-resistant glass sphere which houses 31 PhotoMultiplier Tubes (PMTs), includes the acquisition electronics [6, 7, 8, 9, 10], whose main acquisition board is the Central Logic Board (CLB) [11, 12], supplied by its auxiliary power board [13]. The DOMs are distributed along vertical lines, called Detection Units (DUs), each hosting 18 DOMs. At the bottom of the DU, a base module is installed which provides power to the DU as well as communication. In ARCA, the vertical spacing between DOMs is 36 m, while in ORCA it is 9 m. A thin backbone with fiber optics for communication and copper wires for power, runs along the full DU. The DUs are anchored to the seafloor, with a spacing of about 90 m in ARCA, and 20 m in ORCA, in a regular lattice organized in Building Blocks composed of 115 DUs. Specific instrumentation is installed in the so-called Calibration Unit, of which a few units will be deployed. In order to facilitate and speed up the deployment of the DUs, a custom system has been developed by KM3NeT. The DU is rolled up in a small, re-usable spherical launching vehicle, and, once deployed in the seafloor, the string unfurls to its full length with the buoyant launching vehicle rolling up to the surface [14]. Currently, almost 40 DUs have been deployed and are taking data [15]. The PMT signals are converted into Low Voltage Differential Signals (LVDS) [1] by the PMT base boards. The LVDS duration is equal to the time the PMT signal exceeds a preconfigured threshold, called Time over Threshold (ToT). The LVDS signals are conducted to the CLB by an aggregation board called Signal Collection Board (SCB). The resolution of the reconstructed neutrino trajectory in the detector depends on the accurate measurement of the arrival time of the light on the optical sensors as well as the precise determination of the position of the sensors. Precision of one nanosecond on the arrival time and better than 20 cm on the position of the light sensors is mandatory in order to achieve the required reconstruction performance in the detector. Therefore, time and position calibration of the telescopes are critical [6]. The synchronization of the DOM is performed in the Field Programmable Gate Array (FPGA) of the CLB by means of the White Rabbit protocol [16, 17], which allows for data communication and synchronization using the 1 Gbps optical link available. The identification number of the temperature sensor controlled by the White Rabbit protocol is used to generate a unique Medium Access Control (MAC). The optical network architecture is based on a broadcast optical downlink, while the uplinks are independent communication channels [18]. The optical network architecture of the telescope has evolved to point-to-point connections, requiring a modification of the embedded software. The principle behind the acquisition in KM3NeT is the all-data-to-shore concept, which follows the same principle as in the ANTARES experiment [19]. Through this concept, all PMT signals are sent to the control station on the shore, where the triggering process is carried out on a processor farm. The arrival time and the ToT of the LVDS signals coming from the PMT bases are determined, with one nanosecond resolution, by the Time to Digital Converters (TDCs) [20] implemented in the FPGA of the CLB. The acquisition data is organized in time frames, which are denominated timeslices, with a default duration of 100 ms. The TDCs timestamp the arrival time with the number of nanoseconds after the start of the timeslice. In a later stage, a gateware state machine [21] packs the TDC data into *jumbo* frames (Ethernet frames with more than 1500 bytes of payload, the limit set by the IEEE 802.3 standard), relating the timestamp to the UTC time of the start of timeslice. Once completed, the *jumbo* frames are routed to the specific port of the IPMUX (IP Multiplexer), from where they are sent to the shore station via the White Rabbit PTP Core. The detectors are conducted by the Detector Manager [22]. This system implements the general Control Unit state machine of the detector, which is responsible for setting the input parameters for all DOMs and drives the embedded-software replica state machine managing the DOMs.

[1] https://www.t10.org/ftp/t10/document.95/95-268r0.pdf

The embedded software provides the framework to operate the acquisition system at the DOM in synchronization with all other DOMs. The first generation of the KM3NeT firmware was in development since 2012 and was deployed at the beginning of 2016. Since 2019, the Next Generation (NG) Firmware is under development, implementing many improvements with respect to the first generation, both in project semantics and program architecture.

The scope and goals of the embedded software are presented in Section 2. The hardware environment where the embedded software runs in KM3NeT is introduced in Section 3. A description of the CLB firmware and the main processor where the embedded software runs is given in Sections 4 and 5. Section 6 is dedicated to the architecture of the embedded software. The kernel and the hardware abstraction layer are presented in Section 7, while the network stack is presented in Section 8. The application implementation is detailed in Section 9, while the conclusions are discussed in Section 10.

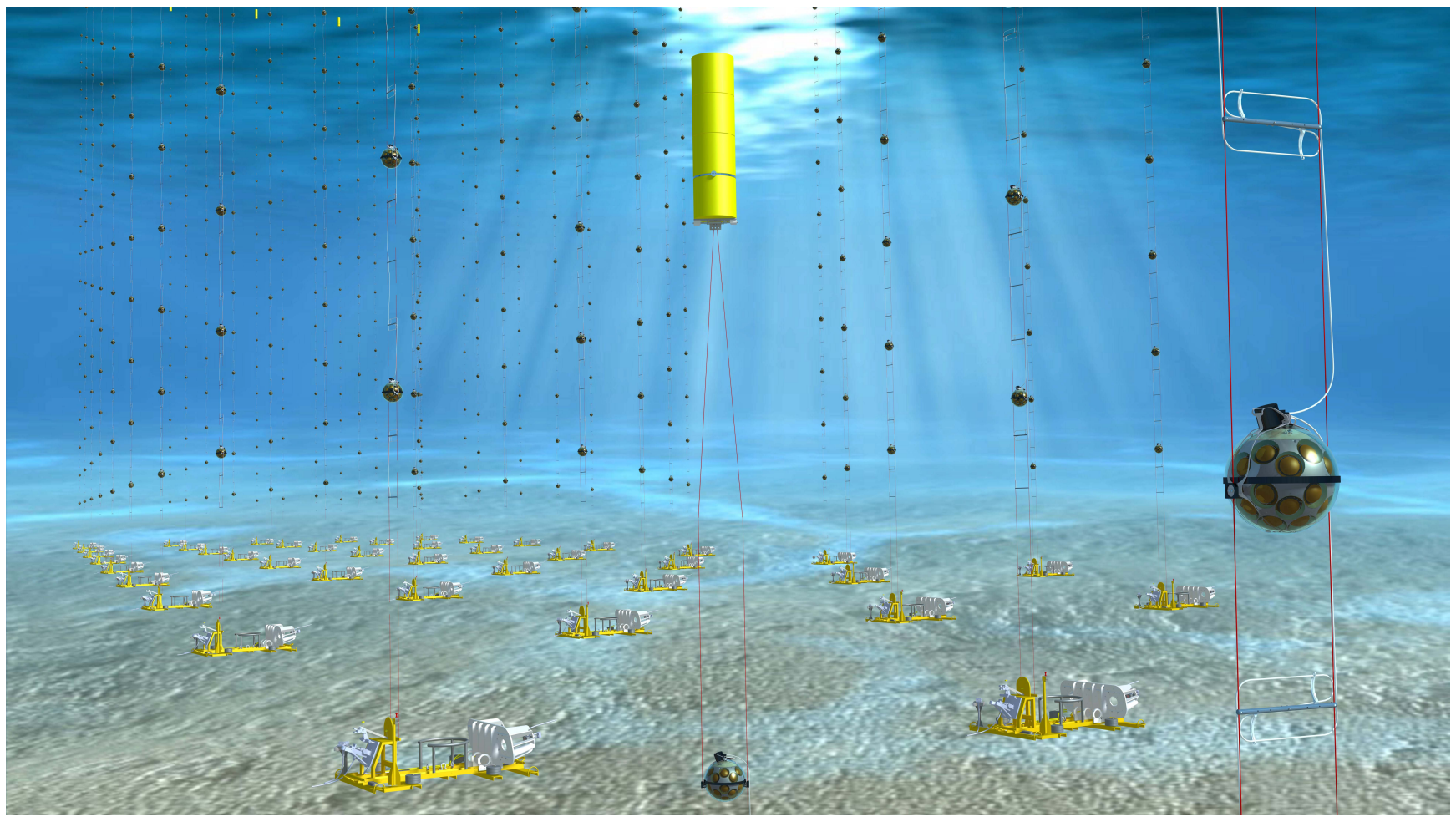

Figure 1: Artist view of KM3NeT. The illustration is not to scale. The sunlight at the bottom of the sea is only for artistic purposes as it will not reach the depths at which the KM3NeT telescopes are installed.

## 2. Scope and goals

The FPGA contains two embedded processors, one LatticeMico32 (LM32)[2] that incorporates the White Rabbit core and a second LM32 added in the KM3NeT logic. The White Rabbit LM32 software has been developed by the White Rabbit Collaboration while the KM3NeT Collaboration has developed the embedded software adapted to the detector network topology. The second LM32 controls the DOM. The main tasks performed by embedded software are:

- initialize, control, and monitor the hardware;
- execute the commands sent by the shore station;

[2]https://www.latticesemi.com/en/Products/DesignSoftwareAndIP/IntellectualProperty/IPCore/IPCores02/La

- align the execution of the program with the Detector Manager Control Unit;
- send diagnostic information back to shore;
- apply firmware updates.

The embedded software will be used in more than 6000 nodes. The KM3NeT telescopes are a heterogeneous detector and there is not a single firmware version for all its modules. The firmware comprises three main different aspects that have to be taken into account:

1. The hardware-specific FPGA bit-file for the different versions of the CLB (v2 or v4). In addition to the version modifications, there are minor hardware modifications in the sensors that should be taken into account during operation.
2. The White Rabbit embedded software for either a KM3NeT custom version or standard White Rabbit.
3. The CLB application software can be either DOM, DU-Base, Calibration Unit-Base or Golden. The last one is a fail-safe image to be used only at start-up. The DU-Base software is a modification to allow the use of the CLB in the Base module of the DUs, where it acts as the controller node of the DU. In a similar way, the CLB is used in the Calibration Unit. A flavor of the CLB software is used to control and monitor the status of the CU.

The embedded software has to be able to deal with future modifications on the hardware, changes in the network architecture and new applications, while providing a global functionality to the whole detector in a seamless way. One of the main constraints in KM3NeT is that the detector is not accessible for maintenance. The embedded software has been designed to provide a set of tools with the sufficient level of reliability to be able to recover hung-up nodes and with enough flexibility to diagnose non-functional elements. Power and cost budget limitations have been other criteria for the design of the embedded software and the hardware resources to operate it.

## 3. KM3NeT Acquisition Electronics

In this section, the different acquisition boards which are controlled by the KM3NeT embedded software are detailed.

*3.1. PMT base board*

The PMT base board is responsible for both the generation of the high voltage required by the PMT and the digitization of the PMT signals [23, 24]. Before being digitized, the PMT signal is amplified in a preamplifier. One of the main components of the PMT base board is a comparator, which provides a high-level logic signal when the PMT output exceeds the comparator limit, which is set via Inter-Integrated Circuit ($I^2C$)[3] by the embedded software. The High Voltage (HV), which is remotely configured via $I^2C$, is generated independently on each of the PMT base boards. This allows the individual gain of the PMTs to be adjusted to equalize the response to photons and provide a ToT value of around 26 ns for the detection of a single photoelectron. The HV value can be remotely adjusted in a range from $-1500$ to $-700$ V. The output of the voltage multiplier circuit is used to power the PMT dynodes.

*3.2. Signal Collection Board*

The Signal Collection Board (SCB) is the board that collects the LVDS signals coming from the PMT base boards, and conducts them to the TDCs implemented in the CLB. The SCB also transfers the $I^2C$ commands from the CLB to the PMT base boards to monitor and control the PMTs. An $I^2C$ controllable complex programmable logic device reads the current sensors and can disable the digital clock to eliminate possible interferences on the PMT signals. The acoustic sensor is also connected to the CLB through one of the SCB.

*3.3. Power Board*

The power board provides power to the entire DOM, including PMTs, acquisition electronics and instrumentation. The DOM is powered by an external 12 V that is input to power board inside the DOM. Different regulated voltages (1, 1.8, 2.5, two 3.3, and 5 V) are generated from this input using DC/DC converters. In addition, the Power Board provides another output, configurable via $I^2C$, which can be set from 0 to 30 V the nanobeacon [25], a time calibration device housed in the DOM. One of the functions of the power board is to provide the start-up sequence of the FPGA voltages. For this purpose, the Power Board incorporates a sequencer that provides the desired voltage sequence. Another feature includes a hysteresis loop that prevents instabilities during start-up. The power board regulators are activated only when the input voltage exceeds 11 V, while they are deactivated when the input value falls below 9.5 V. This prevents fluctuations in the power board regulators.

[3]https://i2c.info/i2c-bus-specification

### *3.4. Central Logic Board*

The Central Logic Board (CLB) is the main electronic board of the KM3NeT acquisition system. The latest version of the CLB is shown in Fig. 2. The main component of the CLB is an FPGA from the Xilinx Kintex-7 family, chosen for its relatively low power consumption, less than 4 W in normal operation, and for being cost effective in relation to its speed. Other relevant components are a flash memory, which communicates via Serial Peripheral Interface (SPI) [4] with the FPGA and stores four of the FPGA images together with the CLB configuration parameters; voltage controlled oscillators, which provide the clock signals needed for the White Rabbit protocol; and two "press-fit" connectors, which provide a solid mechanical and electrical connection between the CLB and the SCBs. The PMT base generates LVDS signals after processing the electrical pulses from the PMT. The SCBs receive these signals and forward them to the CLB, where they are digitized (i.e. timestamped) by the TDCs implemented in the FPGA with a resolution of 1 ns. After collecting timestamps of all PMT base LVDS inputs, the data acquired by the TDCs are sent to the shore control station for further processing and storage. The CLB also incorporates a compass and an inclinometer, three temperature sensors, and a humidity sensor. In addition, it provides a connection for the nanobeacon and for the piezo acoustic sensor. The control of the CLB is achieved through one of the processors embedded in the FPGA programmable logic.

## 4. CLB firmware

The firmware runs on the CLB FPGA and its main components are:

- two embedded LM32 processors: one inside the White Rabbit PTP core that implements the White Rabbit protocol; the second one controls and monitors the CLB;

[4] https://www.analog.com/en/analog-dialogue/articles/introduction-to-spi-interface.html

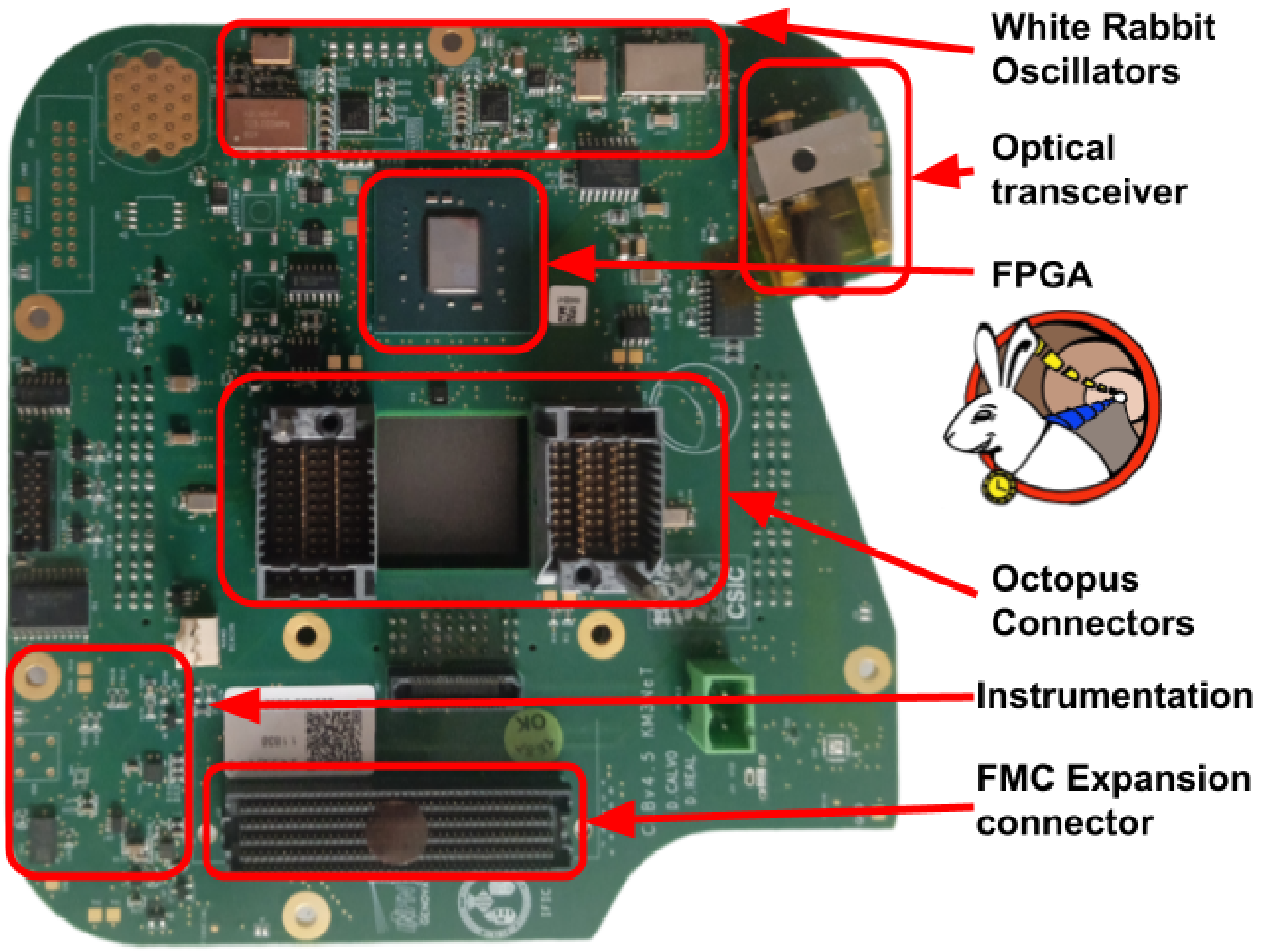

Figure 2: The latest version of the CLB(v4). The embedded software runs on the FPGA. The most important elements of the CLB, apart from the FPGA, are the optical transceiver, the SCB connectors, the White Rabbit oscillator system, and the instrumentation.

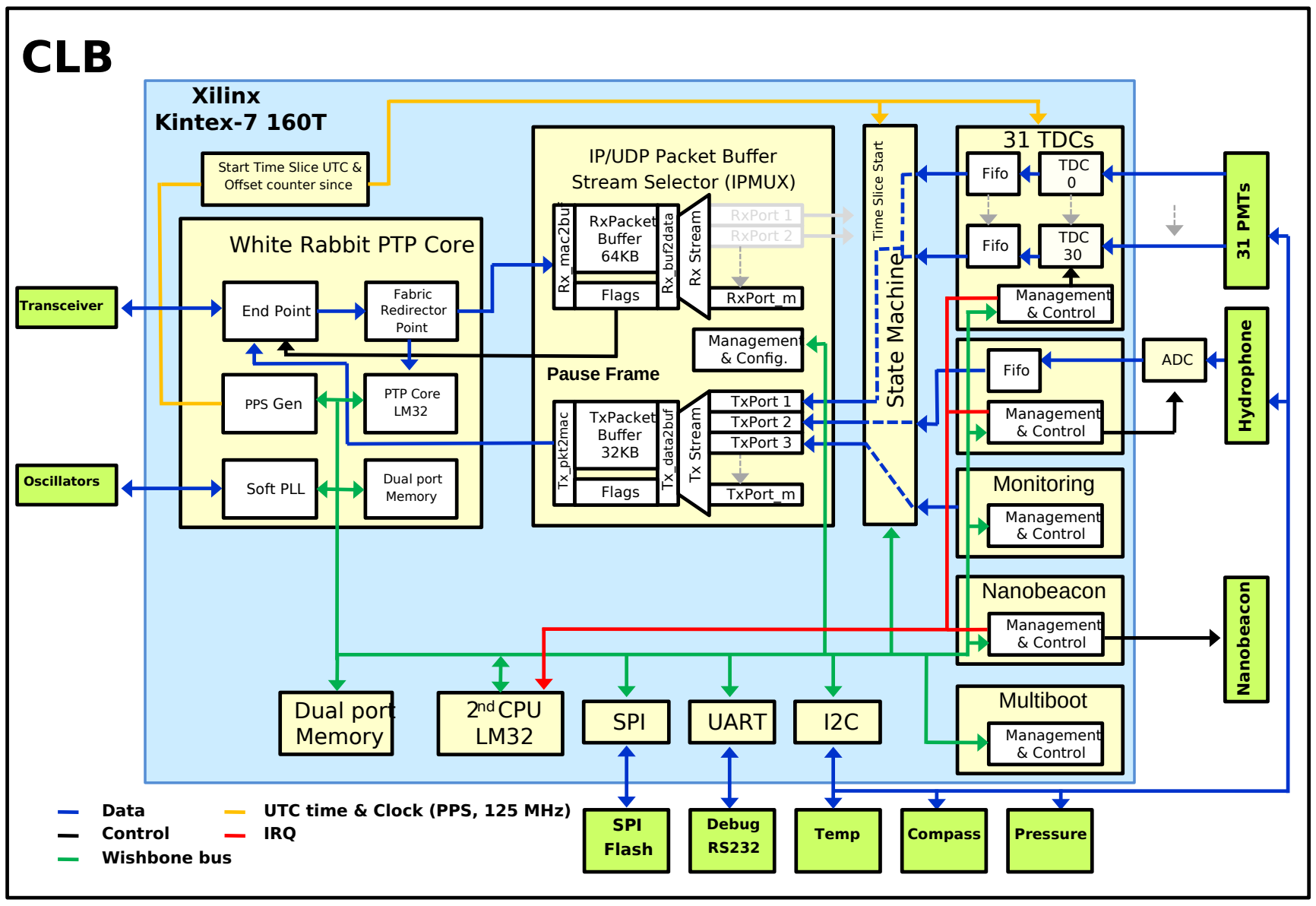

Figure 3: Block diagram of the CLB. Optics, acoustics, instrumentation, front-end firmware and all the interfaces are shown.

- the TDCs, which digitize the PMT signals that arrive at the CLB;
- the IPMUX, which collects the data from the TDCs and the monitoring data generated by the LM32, and sends them via Ethernet to the shore control station;
- the multiboot core, which allows the secure remote configuration of the FPGA firmware.

The architecture of the firmware is displayed in Fig. 3.

## 5. LatticeMico32 soft processor

A soft processor was chosen for the control of the CLB. The two main reasons for this choice are the possibility to integrate it tightly with the acquisition firmware, and the increased reliability. The decrease of speed with regard to physical processors was not a critical point.

Since the WR collaboration selected the LM32 [26] to be implemented in their White Rabbit PTP core, it was decided to select the same processor for implementing KM3NeT specific control and monitoring. The LM32 IP core has been developed by Lattice Semiconductor and it is available under a free IP license. It has a 32-bit Harvard-RISC architecture with a reduced set of instructions and separated instruction and data buses, with a unique address space. The LM32 implements a totally bypassed and interlocked pipeline of six stages, and the arithmetic operations are done register to register. The architecture of the LM32 is shown in Fig. 4. Using the same model of soft processor simplifies both code development and maintenance. The LM32 is open source, it can be ported to FPGAs of different vendors and it is well documented with an existing tool chain. It also allows JTAG access, which is used in KM3NeT to provide debugging support and communications with a flexible bus, such as Wishbone [5]. Another advantage, critical in KM3NeT, is the use of few FPGA resources. An LM32 uses around 2000 logic cells of the XC7K160T FPGA, about 1.3% of the total logic cells [27].

### *5.1. Wishbone Bus Protocol*

The LM32 uses the Wishbone Bus protocol [28]. The Wishbone Bus definition has been done by the OpenCores organization on an open-source basis, being B4 the latest version released and the one used in KM3NeT. Wishbone Bus provides a robust and portable bus standard, allowing for the connectivity with various IP cores. It connects the LM32 with its peripherals. In particular, two Cross Bar Switches (CBS) are used. The first one allows for connecting with the White Rabbit PTP core. The dual memory port and the second CBS give access to all the peripherals and all the acquisition IP cores. The CBS contains a parallel Wishbone Bus used by the Message Signalled Interrupt system to trigger interrupts or exchange short messages. An example of the Hardware Description Language (HDL) implementation of Wishbone Bus is found in Listing 1 and Listing 2 where the slave HDL code for the Wishbone register access and the register map, are presented.

### *5.2. Peripherals and acquisition IP cores*

The LM32 controls a set of peripherals and acquisition IP cores. In order to do so, a map of addresses is generated. The peripherals and IP cores connected to the LM32 are the following:

[5]https://cdn.opencores.org/downloads/wbspec_b4.pdf

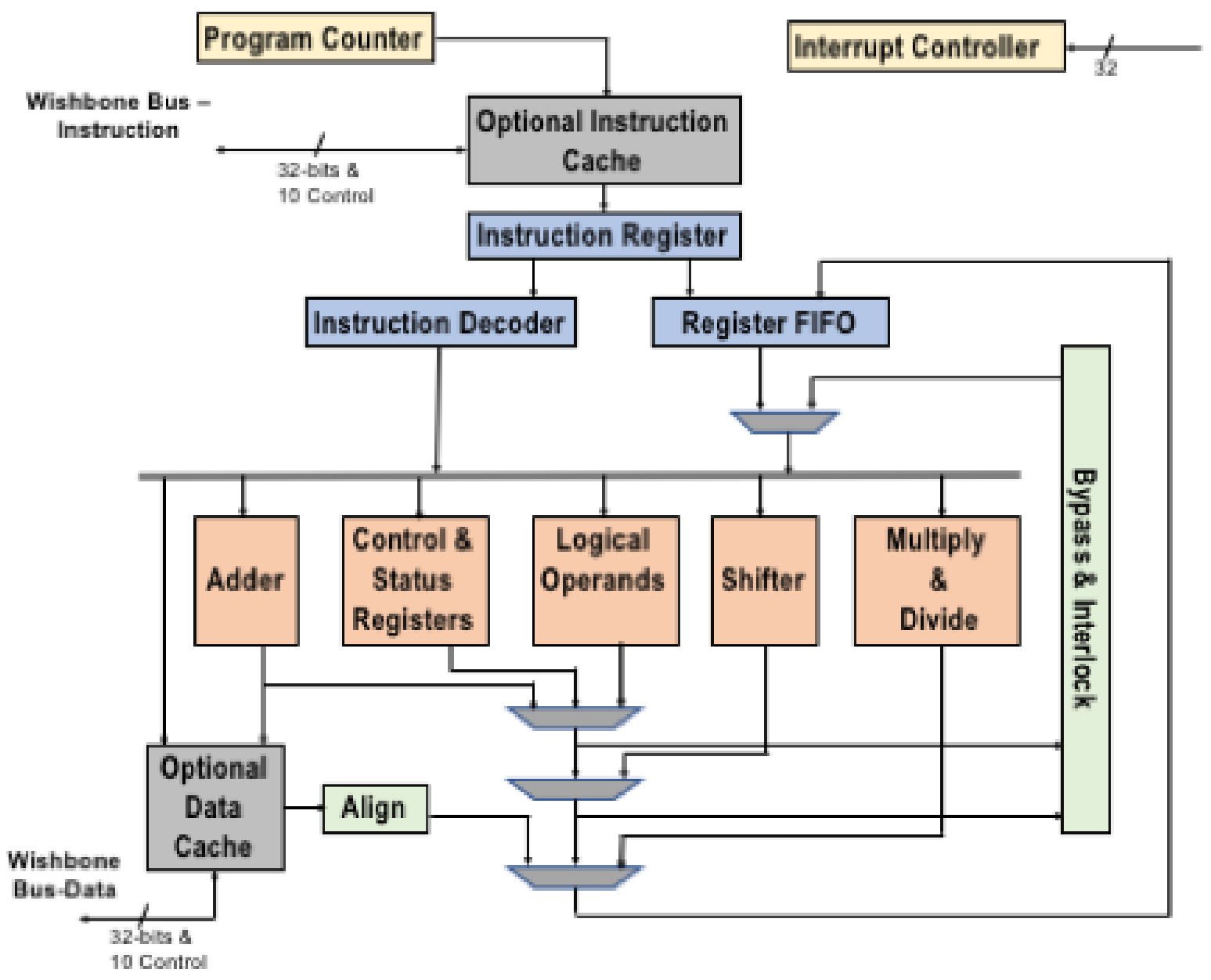

Figure 4: Architecture of the LM32 soft processor where the embedded software runs.

- General Purpose Input Output (GPIO): 16 GPIOs are available for various tasks, including enabling power rails for sensors and managing a watchdog. While debug boards allow control over LEDs and switches, these components are not included in mass-production.
- Timer: A timer is available for the peripherals. Several counting modes are available as well as a prescaler.
- Universal Asynchronous Receiver-Transmitter (UART): Two UARTs have been included in the acquisition system. Via WB, it is possible to set the baud rate, the number of data bits as the parity.
- Serial Peripheral Interface (SPI): The SPI is used to access the SPI flash memories, where the FPGA images, the configuration parameters and the debug logs are stored.
- $I^2C$: Five $I^2C$ buses are used. The configuration and readout of the sensors housed in the CLB, as well as the configuration of the PMT bases, are done via $I^2C$.

- TDCs: The TDCs sample the signals from the bases of the PMTs. They are implemented, one per PMT, in the FPGA of the CLB. A TDC channel measures both the arrival time of the pulse and the ToT, using the time provided by the White Rabbit core. The core of the TDCs produces 48 bits per event. The eight most significant bits are used for the identification of the PMTs, the next 32 encode the arrival time of the event with respect to the timeslice start and the last 8 bits encode duration. The events are sent to the state machine which also organizes the acquisition of TDCs in timeslices. The FPGA clock system is derived from a 25 MHz quartz oscillator. The clock signal is transferred to a digital Phase-Locked Loop (PLL) that generates the 62.5 MHz system frequency. The White Rabbit protocol adjusts the phase and frequency of the system clock speed from the FPGA to the reference master clock. Finally, the adjusted clock is injected into the FPGA PLL to generate the two 250 MHz clocks, identical but shifted 90 degrees in phase. The input of the TDCs is oversampled to one nanosecond using the rising and falling edges of the two 250 MHz clocks.

- TDC State Machine: The core of the state machine orchestrates the data acquisition of the CLB. It is responsible for generating the periodic start of the timeslice signal. This signal is synchronized to the start of the UTC second and is repeated at the beginning of each period. The acquisition synchronizes to the timeslice start signal and the acquired data are segmented and temporally referenced with respect to it. The state machine is responsible for collecting the acquired data and concatenating them to the UTC time of the timeslice start, called *super time*. Once processed, the acquired data are distributed to the IPMUX. The data is sliced into frames in such a way that it can be packed into jumbo frames. A header is prepared with metadata such as the package identifier and the run number.

- IP/UDP Packet Buffer Stream Selector: The packets created by the state machine are sent to one of the input ports of the IPMUX, which acts as a selector for the packet buffer. The IPMUX has different input ports for several data sources: TDCs, acoustic acquisition, monitoring, and slow control channels from the LM32. The IPMUX adds a UDP header for each data packet from these sources before transferring them

```vhdl
entity tdc_master_top
-- ...
port (
    -- wishbone signals (for reading)
    -- master clock input
    wb_clk_i:in std_logic;
    -- lower address bits
    wb_adr_i:in std_logic_vector(2 downto 0);
    -- Databus input
    wb_dat_o:out std_logic_vector(31 downto 0);
    -- Write enable input
    -- ......
  );
end entity tdc_master_top;
```

Listing 1: Details of the definition of the wishbone register access in the TDC HDL top code.

to the White Rabbit core endpoint, via which they are sent to the shore control station. The data type (optical, acoustic, or monitoring) is determined at the shore station by the port origin. The IPMUX is a Wishbone Bus slave of the LM32, so it can be configured remotely.

- Multiboot: When the FPGA is started, it configures itself by loading the first valid image found while scanning the SPI memory. Up to four images can be saved in the flash memory, reserving enough space for storage of CLB configuration parameters.

- Nanobeacon: The nanobeacon trigger signal is generated in the FPGA. It can be configured to change the period, the number of flashes per timeslice, as well as start time of the trigger. All these parameters are set by the embedded software via Wishbone Bus.

## 6. Software architecture

The embedded software running on the CLB is a C-based bare metal application and thus uses no real-time embedded operating system. The embedded software application is loaded into the RAM of the KM3NeT-specific

```
  begin
    if (wb_clk_i'event and wb_clk_i='1') then
      case wb_adr_i is
        --enable channels
        when "001"  => wb_dat_o <= rtdc1;
        --Veto value
        when "010"  => wb_dat_o <= rtdc2;
        --enable veto
        when "011"  => wb_dat_o <= rtdc3;
        --enable multihit
        when "100"  => wb_dat_o <= rtdc4;
        --almost_full_offset
        when "101"  => wb_dat_o <= rtdc5;
        --min ToT
        when "110"  => wb_dat_o <= rtdc6;
        -- ....
      end case;
    end if;
  end process assign_dato;
```

Listing 2: Details of the register mapping. Write access case.

LM32 at FPGA configuration. The application requires less than 256 kB of RAM, including heap and stack. The software architecture consists of two top-level layers: system software and application. The system software layer is the same for each kind of firmware image. This layer contains OS-like features, such as a simple cooperative multitasking scheduler, a firmware update unit, various peripheral drivers, a UDP-based network stack, and support utilities for logging and error handling. The application layer contains a software state machine, slow control command handling, debug shell command implementation, and the KM3NeT core application code composed of a number of subsystems.

The state machine defines the state of the application and drives the application. Implementing the application software as a state machine ensures that the program will always be in an unambiguous and consistent state. The state machine can be navigated by issuing events over slow control from an external driver such as the Control Unit [22]. Additionally, some events are issued autonomously by the embedded software, such as on error occurrence or during system start-up. Application code can be attached to state machine transitions or to periodic timer events. As previously mentioned, the application code is grouped into subsystems. A subsystem is a unit of code and data responsible for the operation of a specific part of the acquisition electronics. It controls hardware peripherals, through the associated driver abstraction. The following subsystems are defined:

- System: application function not specific to other subsystems.
- Optics: control of PMTs and TDCs (only in DOM firmware).
- Acoustics: acoustic sensor control and Audio Engineering Society (AES) protocol handling.
- Instrumentation: sensor readout, generally over the $I^2C$ bus (temperature, humidity, etc.).
- Networking: IPMUX control and White Rabbit monitoring.
- Base: DU-Base module control (only in DU-Base firmware).
- Calibration Unit Base: CU-Base module control (only in CU-Base firmware).

By registering C-functions to state transitions, a subsystem can control the hardware at specific points in the state-machine graph. For example, the start event moves the state machine from `Ready` to `Running`. In this transition, the data-acquisition hardware is enabled to start data taking. The CLB slow control from remote is implemented by means of a custom protocol, on top of UDP. The slow-control protocol consists of three layers. The highest layer is called the Message layer and binds to C-functions at the application level. Messages have a type, e.g. *retrieve firmware version* or *state-machine event*, but also a class, being either Command, Reply, Event or Error. The combination of type and class specifies the format and interpretation of the message payload. Slow-control messages are the primary method for remote control and have functions for moving the software state machine, requesting the status, and many others. Messages are bundled together at the Message Container Format (MCF) layer, binding multiple messages into a single payload for efficiency. The lowest slow-control layer is Simple Retransmission Protocol (SRP) and is responsible for transmission control. It implements a simple packet-based retransmission scheme where the packet contains an identifier ordinal which must be acknowledged within a specific time window and is otherwise retransmitted.

### *6.1. Image types and multi-stage boot*

The SPI flash contains up to four FPGA images. The first image is the Golden image which should not be changed unless absolutely required. The Golden image is always the first image to load, and its primary purpose is to start the run-time image. For this reason, the Golden image is a small image, containing only the basic necessities for booting the second image, which could either be the DOM, Base, or Calibration Base image. The Golden image can also update the firmware on any of the other locations. The Golden image waits for 30 seconds after obtaining an IP address such that the default boot procedure can be aborted in case of failures on the other images.

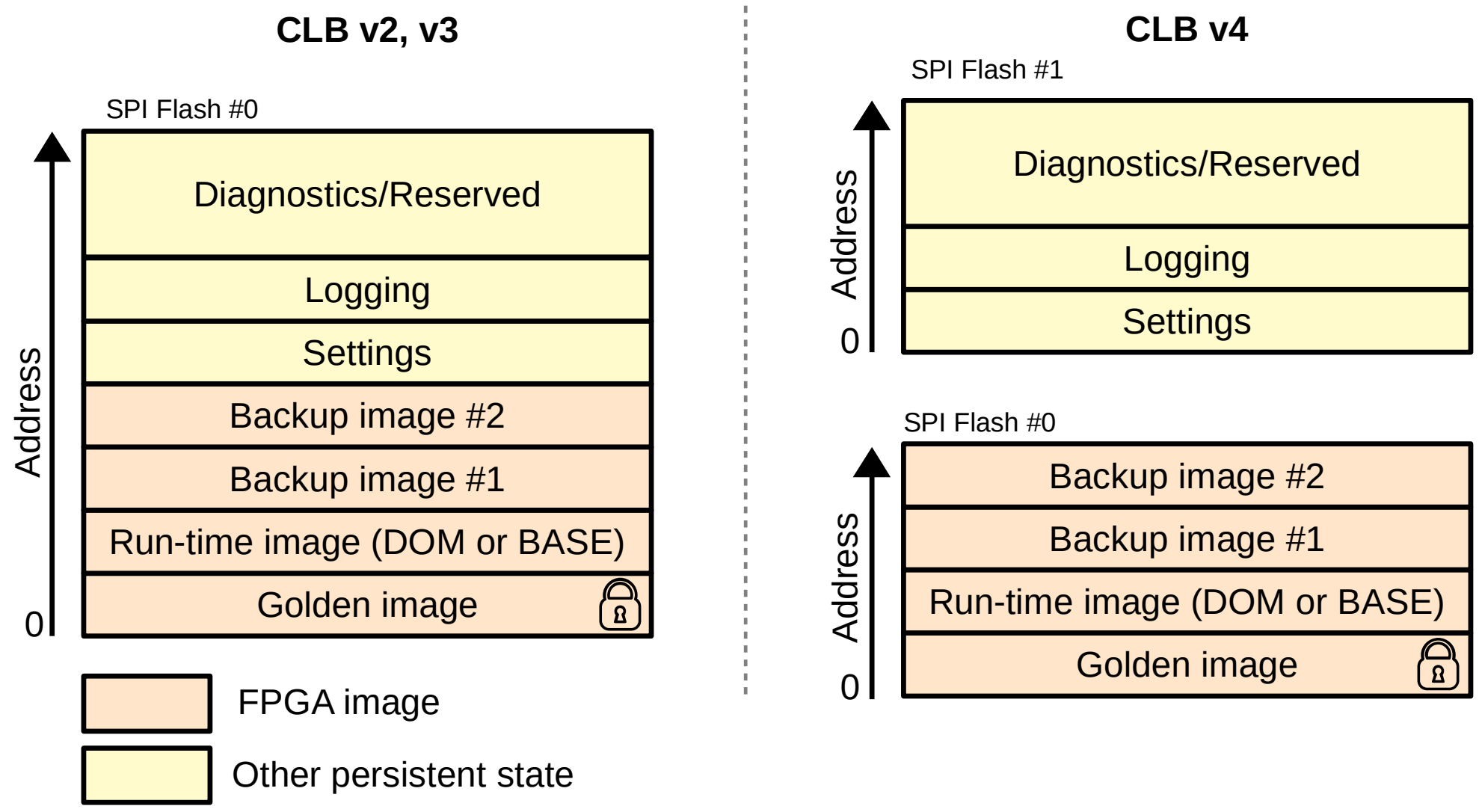

Figure 5: Persistent storage layout, containing both FPGA images and other persistent storage, grouped by CLB hardware version.

While generally the second image is started from the Golden image, the CLB can be configured to automatically start the third or fourth image. The persistent storage layout, including the FPGA images, is shown in Fig. 5.

While CLB v2 [29] contained a single SPI flash chip for storage and images, the CLB v4 [30] features two flash chips for better separation of responsibilities and improved robustness. In addition, the Golden image is protected against accidental overwrite by using the SPI flash block protection features, and requires a password to unlock.

### *6.2. Build environment and procedure*

Project improvements include the usage of modern development methodologies and tools such as control version repositories (Global Information Tracker (GIT) with sub-modules), Continuous Integration/Continuous Deployment (CI/CD) and Docker containers. These tools and methodologies are used for a consistent and reusable build environment.

The main repository contains software, gateware and hardware files. The software directory contains the software sources and build files to generate the software binaries for different applications (DOM, DU-Base, CU-Base and Golden) and hardware versions (CLB v2 and v4). Additionally, it can execute unit tests and build software documentation. A special branch of the White Rabbit PTP core, maintained by KM3NeT, is included as a submodule within the software directory. This branch contains the broadcast version of the White Rabbit PTP core [18, 31]. The gateware directory consists of scripts and sources used to build FPGA images for different hardware versions (CLB v2 and v4). Since the White Rabbit PTP core is a crucial component of the KM3NeT firmware, the White Rabbit gateware GIT repository is included as a submodule, and the HDL files within it are utilized in the gateware design.

The root directory houses the super-build script, responsible for orchestrating the gateware and software sub-builds and merging the resulting binaries into distinct firmware images aligned with specific applications and hardware versions.

CMake is used as the primary build tool. The entire project can be built inside Gitlab-CI/CD, using a KM3NeT-specific Docker container. Docker containers allow for isolating and simplifying the creation of environments that contain all the libraries, dependencies and binaries to execute programs. In KM3NeT, the containers facilitate the process of compiling the LM32 code used by the CLBs.

There are two contexts: First generation, based on the SVN repository with ISE Xilinx tools and LM32 compiler tool chain; Next generation, based on the GIT repository with Vivado tools and LM32 compiler tool chain.

The creation process of the toolset is based on the Makefile to compile recent tool versions. The container can be created in three different ways:

- Ultrafast: using the Docker container image with CentOS 7 and all the default libraries,
- Fast: using the prebuilt toolchain with CentOS 7,
- Complete: personalised toolchain.

*6.3. Layering*

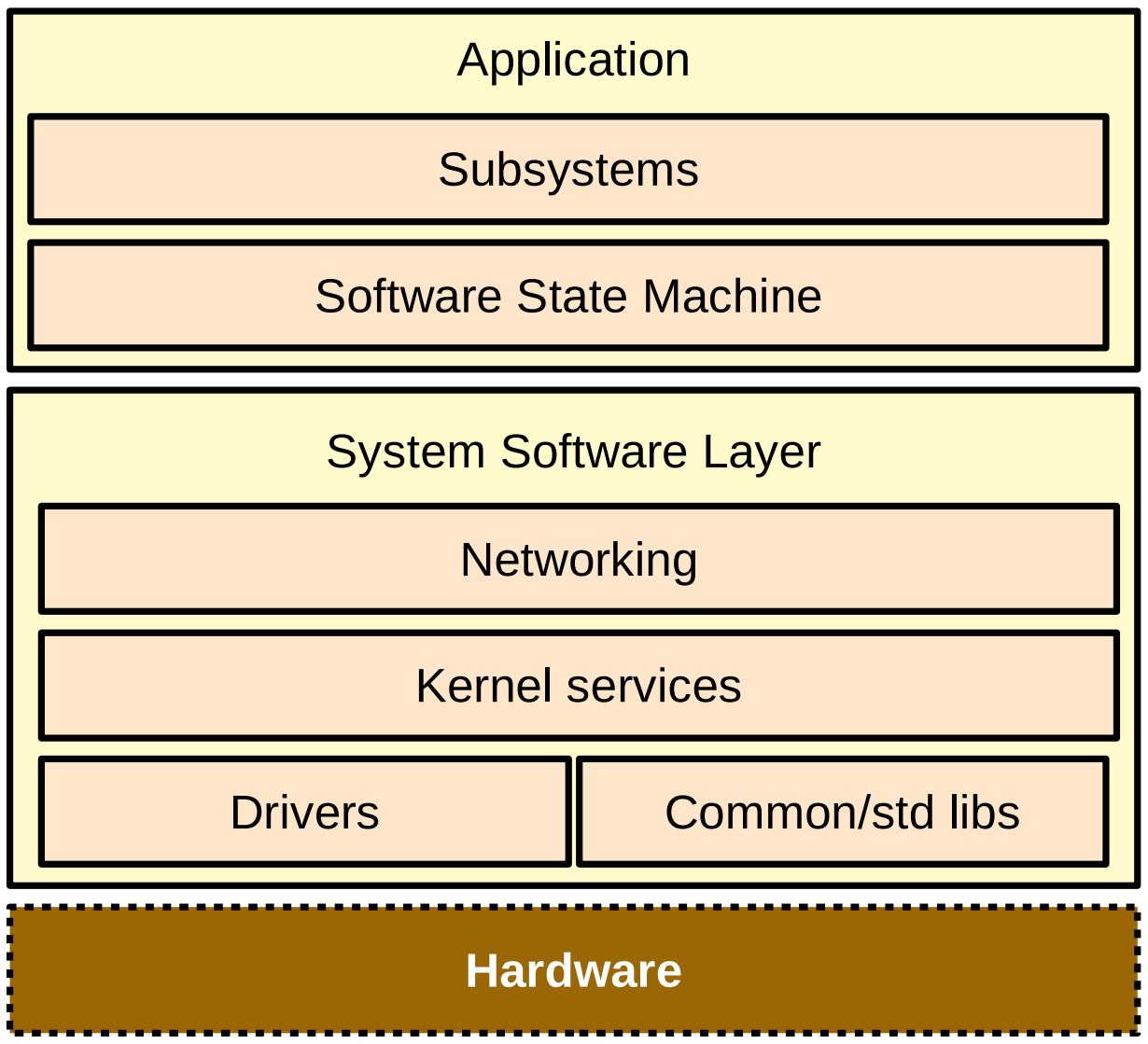

Figure 6: High-level stack of the software operational on the KM3NeT specific LM32. The application and the system software layers are shown.

A high-level stack of the layers of the embedded software running on the KM3NeT-specific processor is presented in Fig. 6. The application layer performs user-specific functions. In the KM3NeT case, this primarily concerns executing commands from the shore station and controlling the detector hardware through the system software layer-provided services. This layer manages functions like the system start-up, hardware control and resource allocation. However, while the system software layer serves to facilitate, it does not define the application itself. It is the responsibility of the application layer to control the system software layer in order to provide specific functionalities. In the next Sections, each of these functions will be explained in more detail.

## 7. Kernel

The kernel is responsible for OS-like functions such as system start-up, scheduling, and resource management. It also contains the hardware abstraction layer including drivers for all controlled peripherals.

### *7.1. Start-up*

The LM32 does not contain an interrupt vector table. Instead, it assumes that interrupt handlers are present at specific offsets in the memory, spaced 8 instruction words apart, starting from address 0. The reset handler is the first interrupt handler and is located at memory address 0, which is also the first instruction that the LM32 executes after a power-up or a reset. A C file contains the interrupt handlers and support code. It also contains code for initializing the C-code variables and jumping to the main function. A second file contains the LM32-specific functions, such as requesting the cycle counter, and disabling/enabling Interrupt Requests (IRQs). In addition, it also contains the IRQ handler code, responsible for handling all IRQs and executing the peripheral-level IRQs. A third file contains the main function, which initializes the system integrity checks, basic system bus controllers, such as UART, $I^2C$ and SPI controllers, and OS services, such as persistent logging and the firmware update system. A call is made to the application-specific initialization before handing over responsibility to the scheduler.

### *7.2. Scheduler*

The KM3NeT embedded software does not use a real-time operating system. Instead, it uses a simple cooperative scheduler. This scheduler can be provided with *tasks* which take the form of a function pointer. The function signature can have no return value and no argument. On registration, the scheduler assigns an identifier to the task, which can be used to refer to it. A task can be scheduled to run on request and/or periodically. On registration, a task may be set as high-priority, indicating that this task should be executed before any normal-priority task, as multiple tasks may be pending for execution at a given time. The scheduler is not preemptive, and once a task is started, it must run to completion, thus a task may stall the scheduler. All tasks must therefore limit their execution time and implement a timeout scheme when waiting for resources. Once the embedded software start-up is completed, the non-returning function `schdExec()` is called which executes the scheduler internal main loop. Besides scheduled tasks, the scheduler can also be provided with idle tasks. These tasks are executed repeatedly as long as there are no scheduled tasks pending for execution. One important idle task is the software integrity check function. This is a best-effort code corruption detection function running a 32-bit Cyclic Redundancy Check on the code and on constants. It also implements a stack-overflow detection through memory markers.

Another important function of the scheduler is to off-load IRQ handling to the main loop. Where possible, an IRQ only incites the scheduler to run a specific task in the main loop. This limits the required critical sections and reduces code complexity. As the scheduler is cooperative, the incited task will need to wait until currently running tasks complete, before the IRQ task can be run. This is different from preemptive schedulers, where the IRQ could trigger a high-priority task which can temporarily suspend the currently running thread. For this reason, IRQ handling tasks are generally put in the high-priority group, allowing as quick as possible handling of the incited task.

### *7.3. Error handling*

Most errors that can occur during run-time are handled using the error module. The convention of fallible functions is to return a Boolean value indicating whether the operation was successful (True) or failed (False). In addition, this function is expected to set a global error code using the `errSet(ERROR_CODE, ...)` pre-processor macro, much like `errno` is used in the POSIX standard [32]. The macro adds the line and unit file to the error for increased traceability, and also allows for additional contextual arguments. For example the code from the $I^2C$ driver responsible for handling the *arbitration lost* condition is shown in Listing 3. This condition occurs when the bus master loses control over the bus. Though the general error is a bus error, indicated by `E_BUS`, additional context is added by specifying the $I^2C$ specific error code `I2C_ERR_ARB_LOST`.

The error module also supports adding context after the error has been generated. In this way, the context can be added at multiple levels. Listing 4 shows a snippet of code from the serial flash driver in which additional context

```
if (dev->CMDSTS & I2C_STS_ARBLOST)
{
  return errSet(E_BUS, "i2c_err", I2C_ERR_ARB_LOST);
}
```

Listing 3: Error handling code for the I$^2$C arbitration lost condition.

is added to an error condition. At the application level, the error including

```
  if (!sfTxRx(dev, dta, dta, 1 + len))
  {
    return errTrace("sf_reg", reg);
  }
```

Listing 4: Adding the serial flash register address to the error generated inside `sfRead`.

the context information is logged, and generally, the application is moved to a safe state in which the error can be retrieved and the application reset. This procedure is discussed in Section 9.2.

### *7.4. Firmware update*

Users can update any of the four firmware images located in the SPI flash remotely, and any of the CLB firmware images has the ability to update the system. To update the Golden image, an additional unlock command is required, removing the hardware protections in place. Note that in most cases updating of the Golden image is not needed, and is discouraged. Furthermore, overwriting any existing image is only allowed if there is at least one more valid image present in the flash memory. If during programming there is a power failure this safe guard ensures at least one loadable image is available as fallback.

Before updating, the firmware-update application checks hardware and application compatibility. To prevent a partially updated image from being read by the FPGA configuration state machine, writing of the image is executed in a specific order. First, the location of the FPGA sync header is cleared. In case a previous FPGA image was present, it will no longer be recognized by the FPGA reconfiguration state machine. Then, the FPGA image is written page by page from back to front. This causes the sync header to be the last page written. By following this approach, power loss during flashing will not result in the configuration state machine reading a partial image, thereby mitigating the risk of rendering the FPGA inoperable. Consequently, the configuration state machine will continue its search until it locates a valid image, which will then be loaded.

### *7.5. Inter-process communication*

The KM3NeT design contains two processor cores, introduced in section 4. In order to facilitate communication between them, a hardware mailbox implementation is present. This mailbox ensures a synchronized communication by implementing a small dual-port RAM memory, mapped to both cores. A protocol is implemented as a memory-mapped C structure defined in a header shared between the two applications, each running on a core.

The C structure consists of a status section, in which the White Rabbit PTP core can write information such as state and sensor data. The second section contains a command/reply framework. For this framework, the KM3NeT core initiates requests to the White Rabbit PTP core by loading a command into the shared structure. The White Rabbit core executes the command and sets a reply into the structure. Finally, it clears the command field, acknowledging the command has been handled.

For instance, this command/reply semantic is employed to modify the auto-negotiation mode. Additionally, it is utilized for wavelength tuning, enabling the adjustment of the transceiver's wavelength.

### *7.6. Hardware abstraction layer*

The platform layer contains a hardware abstraction sub-layer marked *Drivers* in Fig. 6. This layer offers a high-level API for on- and off-FPGA peripherals. On-FPGA peripherals include the DAQ devices such as the TDCs, and AES, but also the SPI and $I^2C$ master bus controllers. Off-FPGA component support includes environmental and compass sensors.

For some functions of the embedded software, there are interfaces known as *modules.* These modules combine multiple low-level drivers into a unit specialized on an aspect of the CLB. For example, the *power* module provides annotated access to the supply rail sensors, controls the voltage to the LED beacon, and toggles the external sensor power rails. The module accesses various $I^2C$ peripherals and GPIO peripherals to perform its function.

## 8. Network stack

KM3NeT contains an in-house developed small-footprint UDP/IP stack, called *ministack*, responsible for handling all the slow control communications. To limit software complexity and memory requirements, no TCP/IP is used. To compensate for the lack of delivery guarantee of UDP, a simple reliability layer use SRP, has been added on top of UDP. This protocol allows for retransmission of lost packets. On top of SRP, the MCF may bundle multiple messages using MSG format for increased protocol efficiency. The three facilitating packet protocol formats are shown in Fig. 7 and are explained in sections 8.3 and 8.4.

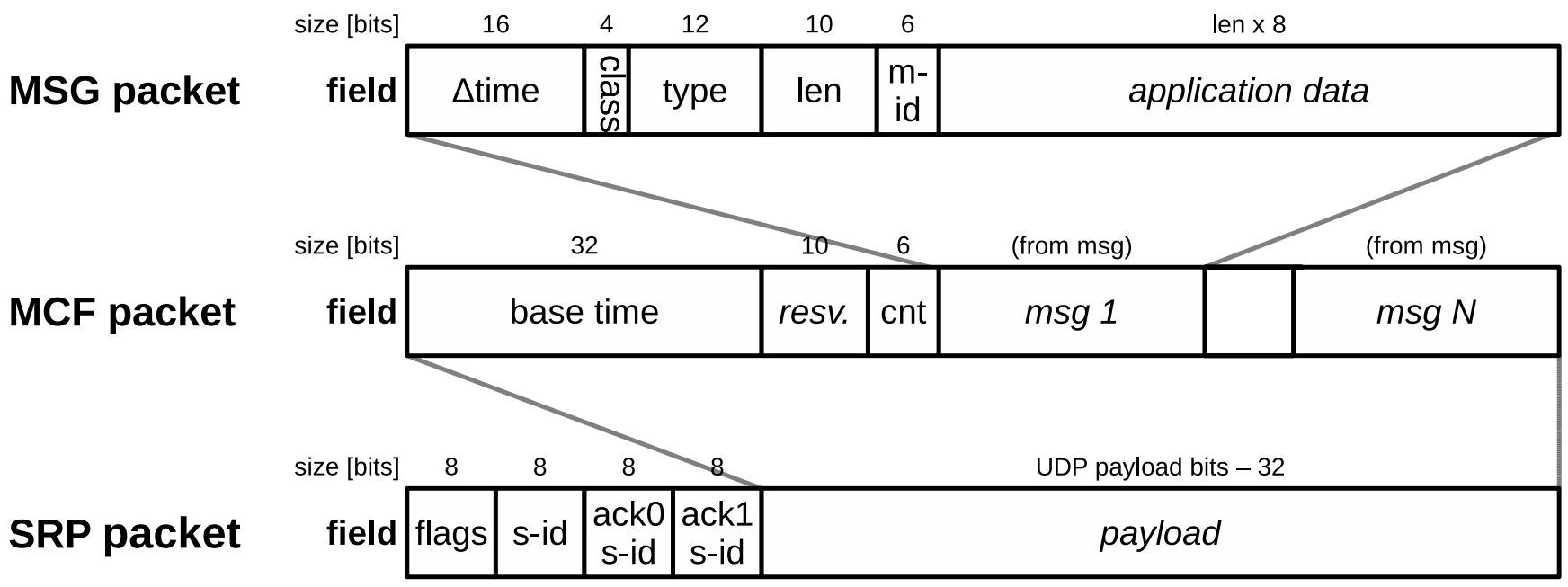

Figure 7: The UDP-based CLB network stack consisting of three facilitating packet types.

### *8.1. White Rabbit MAC/IPMUX Interface*

Incoming Ethernet packets from the detector network are routed through the White Rabbit MAC fabric. A hardware-level packet filter processes each packet and flags it according to a simple program loaded into the filter. A packet may be dropped, routed to the PTP core for timing processing, or sent onward to the IPMUX. The IPMUX then puts the received packet into a software-readable FIFO and provides an interrupt to the software to signal that a packet is ready. Additionally, there is an out-going FIFO used by the software to send out-going network packets.

### *8.2. Ministack*

The ministack is a lightweight UDP stack developed in-house. In addition to its role in parsing and formatting UDP packets to and from the IPMUX, the software also incluses a BOOTP handler for IP acquisition, ARP functionality for MAC address resolution and provisioning, and ICMP support for responding to ping packets. The UDP packet content is forwarded to the higher layer, which in this case is the SRP protocol.

*8.3. Simple retransmission protocol*

SRP implements a simple packet-based re-transmission scheme where each packet contains an identifier ordinal (`s-id`) which must be acknowledged by the remote side by replying with an SRP message having the `s-id` filled in either the `ack0` or `ack1` field (See Fig. 7). This must occur within a 200 ms window and is otherwise re-transmitted. If no acknowledgment is received, the message is sent again with a small delay between each attempt, up to a maximum of 6 times, after which it is deemed lost. An SRP reply may in addition also contain a new SRP message; this is indicated by bits in the `flags` field.

*8.4. Message container format*

The MCF packet bundles multiple application-level messages together but also adds meta-data to each message. When created, each MCF message contains the system up-time in milliseconds in the `base time` field. Up to 64 messages (MSG packets) may be bundled in one MCF packet. Each message contains the creation time (`base time` + `Δtime`), a `class`, `type`, length (`len`) and message-id (`m-id`), as layed-out in Fig. 7. The `class` + `type` identifies the format and meaning of the application-level data. A class encodes the transaction mode of the message which may be `command`, `reply`, `event` or `error`. The type identifies the message function. For example, to retrieve the system build data and revision, the remote side sends a message of class `command` with type `SYS_DATEREV` and no payload. The CLB will reply with a message of class `reply` (when successful) and the same type. This time the payload is the build date and software revision. Only the `error` class has always a fixed encoding regardless of the type, containing the error code and contextual information about the error. The `event` class is a special message which does not expect a reply, and may be broadcast. Each message has a message identifier (`m-id`). This identifier allows tracking of command/reply transactions. The reply to a command must contain the same message identifier. This allows the binding of a reply to an earlier send command, facilitating up to 64 transactions to be in progress at any time for a specific CLB.

## 9. Application implementation

### *9.1. Process variables*

Process variables are a set of remotely-accessible variables for process control. In this case, the process to control is the CLB-embedded application. All variables are defined in a JSON5 file (`variables.json5`). A custom generic template-rendering application, written in python (`jinja2runner.py`), is used to render the variable definition file to an output file in a specific format depending on the provided Jinja2 [33] template as illustrated in Fig. 8.

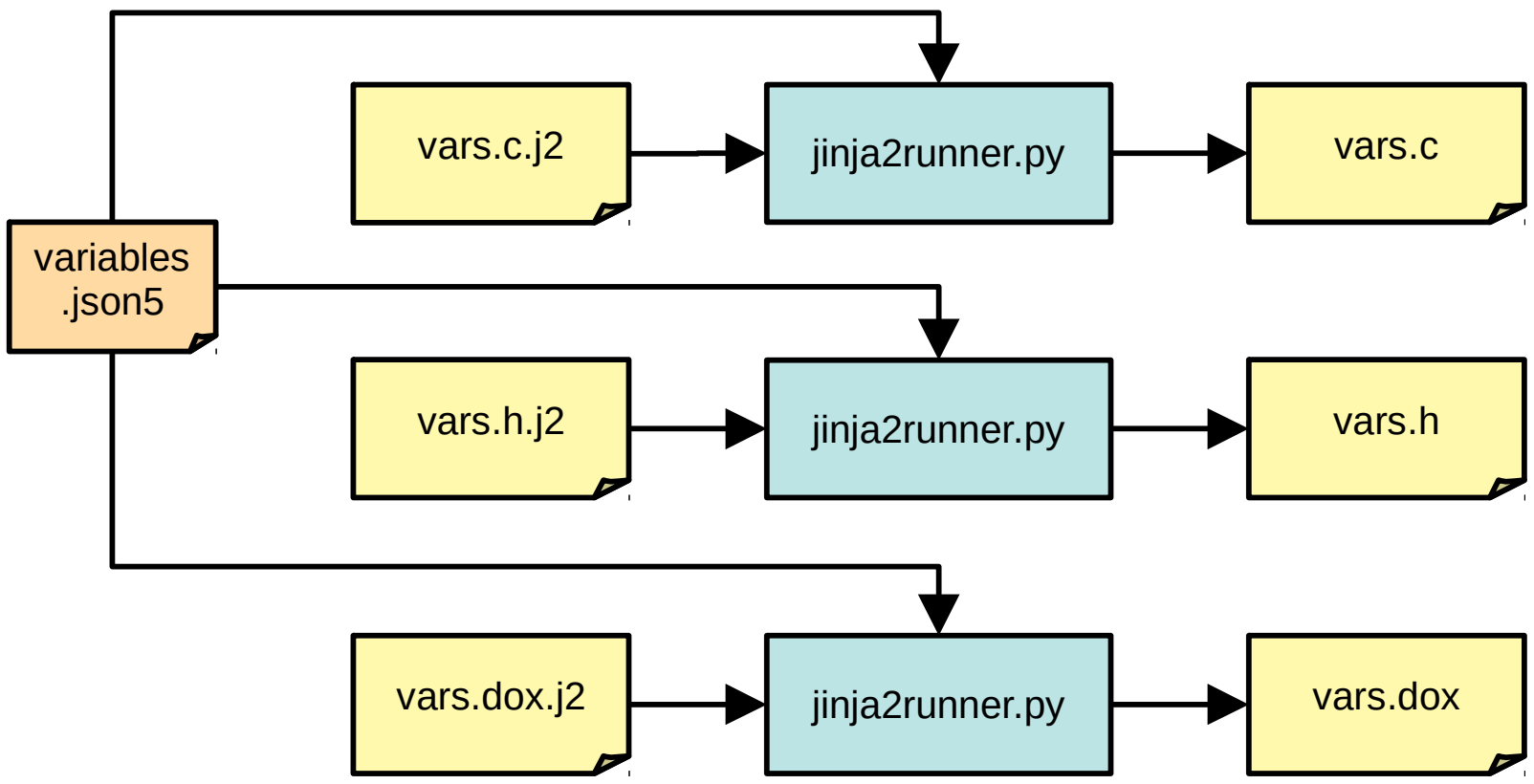

Figure 8: Illustration of the process for generating C-files and documentation files from the source variable definition file.

The `variables.json5` file consists of a list of variable definitions as shown in Listing 5. All variables belonging to a group are loosely related to the subsystem they are part of. A qualified variable name is composed of a short group name and the variable name itself. In the case of Listing 5, the complete naming would be `acs.acou_res` or `ACS_ACOU_RES` depending on the code generation template. Furthermore, each variable has an index `idx` $(1 \ldots 63)$, which becomes part of the variable identification number. Each variable has external access options, either read-write, read-only or write-only (coded respectively as, `RW`, `R` or `W`), and a type. Integer variable types are a combination of either `'I'`, for signed or `'U'`, for unsigned, and the number of

```
{
  group : "ACS",
  name  : "ACOU_RES",
  idx   : 3,
  type  : "U8",
  access: "RW",
  tags  : "C",
  enum  : {
    "12_BITS": { value: 0, desc: "12 bit resolution" },
    "16_BITS": { value: 1, desc: "16 bit resolution" },
    "24_BITS": { value: 2, desc: "24 bit resolution (default)"
    }
  },
  reset : "24_BITS",
  desc  : "Acoustic data resolution"
}
```

Listing 5: Snippet of the `variables.json5` file defining the acoustic resolution options. The template rendering application uses this file to generate the C files needed by the embedded software.

bits (8, 16, 32 or 64). Other supported types are `bool` for Booleans and `f32` for IEEE-754 32-bit floating point values. A `tags` string contains a number of single characters, each indicating an additional option: `C` for configurable, `D` for deprecated and `F` for fallible. A variable marked as configurable can not be modified once the system has passed the `configure` event. Deprecated marks a variable as no longer in use, though generally backward compatibility is supported for a number of versions. Fallible variables are variables which require an explicit validity tag, as their value may not be valid. Generally, this flag is used for external sensors for which readout could fail. The last option for each variable is the `count` (not shown in Listing 5) which indicates the number of elements in this variable. When count $> 0$ this creates an array instead of a scalar. Only single-dimensional arrays are supported.

Optionally, an enumeration can be added to a variable using the `enum` keyword, indicating the possible named values for this variable. The `reset` keyword indicates the value of the variable upon reset, and `desc` adds a description to the variable. During the code generation, the description is generally added as a comment. For each variable, a unique value is derived from some of the elements of the variable definition (Fig. 9). The advantage of this approach is that the application can determine the properties, such size and type, of any variable, even through the variable may be unknown. The size $s$ can be determined for variable ID $v_{\mathrm{id}}$ using

$$s(v_{\mathrm{id}}) = 2^{v_{\mathrm{id}}[17..16]}(v_{\mathrm{id}}[11..0] + 1) \tag{1}$$

where $v_{\mathrm{id}}[m..n]$ provides the unsigned integer value of bits $m..n$ inside the provided variable ID, shifted such that $n$ is the least significant bit. By convention the least significant two bits of the type field, when raised to the power of 2, provide the size of the component type in bytes. When a process variable describes an array its maximum size is 4096 (count = 4095) and the minimum is 1 (count = 0). In the latter case, the process variable is deemed to a scalar, thus arrays of a size less than 2 can not be encoded.

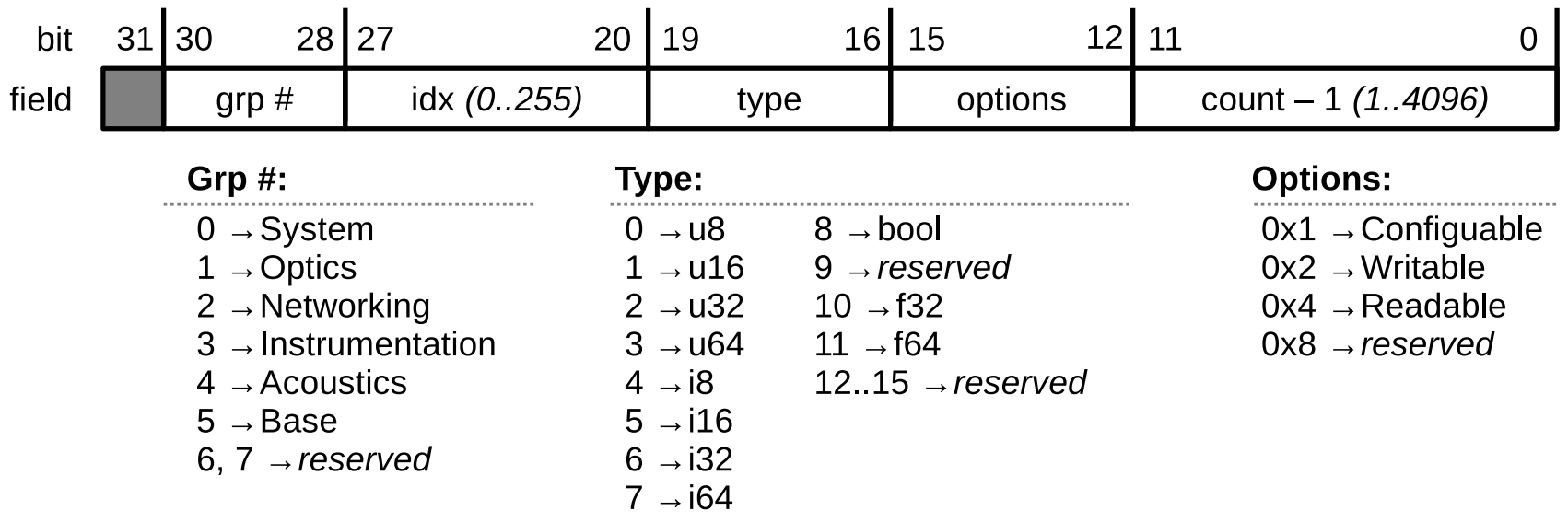

Figure 9: Variable ID composition, including a summary of the sub-fields inside the identifier. The three sub-fields are group, type and option. The group sub-field refers to the system to which the variable belongs. The type sub-field defines the data type and size of the variable. The option sub-field specifies the access right to the variable. The count sub-field specifies the number of scalar element in the variable.

The `variables.json5` file is exported in a separate `shared` repository for sharing between various projects. The file is also used to generate a Doxygen reference, included in the firmware documentation.

For the embedded software, `vars.c` and `vars.h` are generated converting the JSON description into structures for direct access inside the embedded software. An example can be seen in Listing 6 where the generated structure defining the acoustics subsystem is shown, part of `vars.h`.

For all fallible variables, an additional structure is generated to which

```c
// ----------------------------------------
// Definitions for subsystem Acoustics
// ----------------------------------------
#define ACS_ACOU_CHAN          0x40207000

//!< Enable both acoustic channels
#define ACS_ACOU_CHAN_BOTH     0
//!< Enable only channel 1 (default)
#define ACS_ACOU_CHAN_ONE      1
//!< Enable only channel 2
#define ACS_ACOU_CHAN_TWO      2


#define ACS_ACOU_RES           0x40307000

//!< 12 bit resolution
#define ACS_ACOU_RES_12_BITS   0
//!< 16 bit resolution
#define ACS_ACOU_RES_16_BITS   1
//!< 24 bit resolution (default)
#define ACS_ACOU_RES_24_BITS   2

//! Structure defining all process variables for subsystem
↪  Acoustics.
struct acs_s
{
  /** Acoustics channel config */
  uint8_t    acou_chan;
  /** Acoustic data resolution */
  uint8_t    acou_res;
};

typedef struct acs_s acs_t;

//! Provides access to all process variables of subsystem
↪  Acoustics.
extern acs_t acs;

```

Listing 6: Snippet of the generated code of `vars.h` defining the acoustic structure. This code is automatically generated by the template render application.

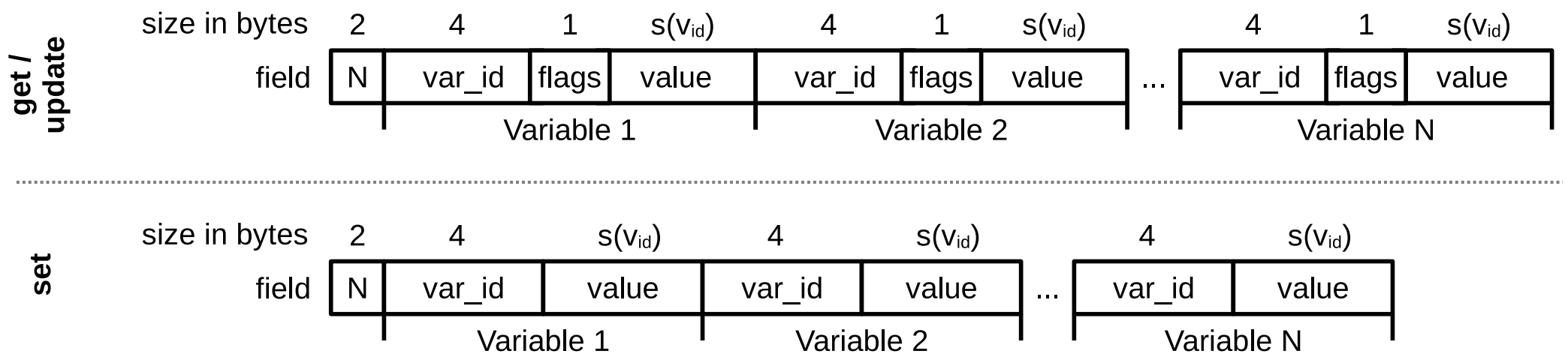

Figure 10: Format of variables returned either when using `MSG_CLB_GET_VARS2` command or inside the `EVT_CLB_UPDATE_VARS2` event (top), and when set using `MSG_CLB_SET_VARS2` (bottom).

the validity of the variable can be set. On reset, the validity of all fallible variables is set to false.

In the embedded code, process variables can be accessed using their structures. For example, `acs.acou_chan` provides access to the acoustic channel settings and can be set with plain C code, e.g. `acs.acou_chan = ACS_ACOU_CHAN_TWO`. These structures are always readable and writable by the embedded software.

*9.1.1. Remote access and monitoring*

In addition to these structures accessible from the embedded code, the code generation also creates meta-data and a look-up table for remote access. The meta-data are used by the `access.c/access.h` to provide high-level functions and allow variable introspection. Together with the `buffer` module, it allows for the serialization and deserialization of any process variable.

A remote client can read, write and subscribe to variables using the `MSG_CLB_GET_VARS2`, `MSG_CLB_SET_VARS2` and `MSG_CLB_SUB_VARSRATE2` message types respectively.

For some variables, such as sensor data, it is important to receive frequent updates. For this reason, there is a subscription system, allowing periodic transmission of a subset of the process variables. When subscribed, each variable is sent using the `EVT_CLB_UPDATE_VARS2` event with the configured interval, spanning from 1 to 127 s with a granularity of one second.

Whether variables are queried or sent using the subscription system, the format of the sent data is the same as shown in Fig. 10. The variable list contains a sequence of ID, flags, and value. The flags byte is used to communicate the valid state. For variables which are not fallible, the flags are always set to valid.

*9.2. Software state machine.*

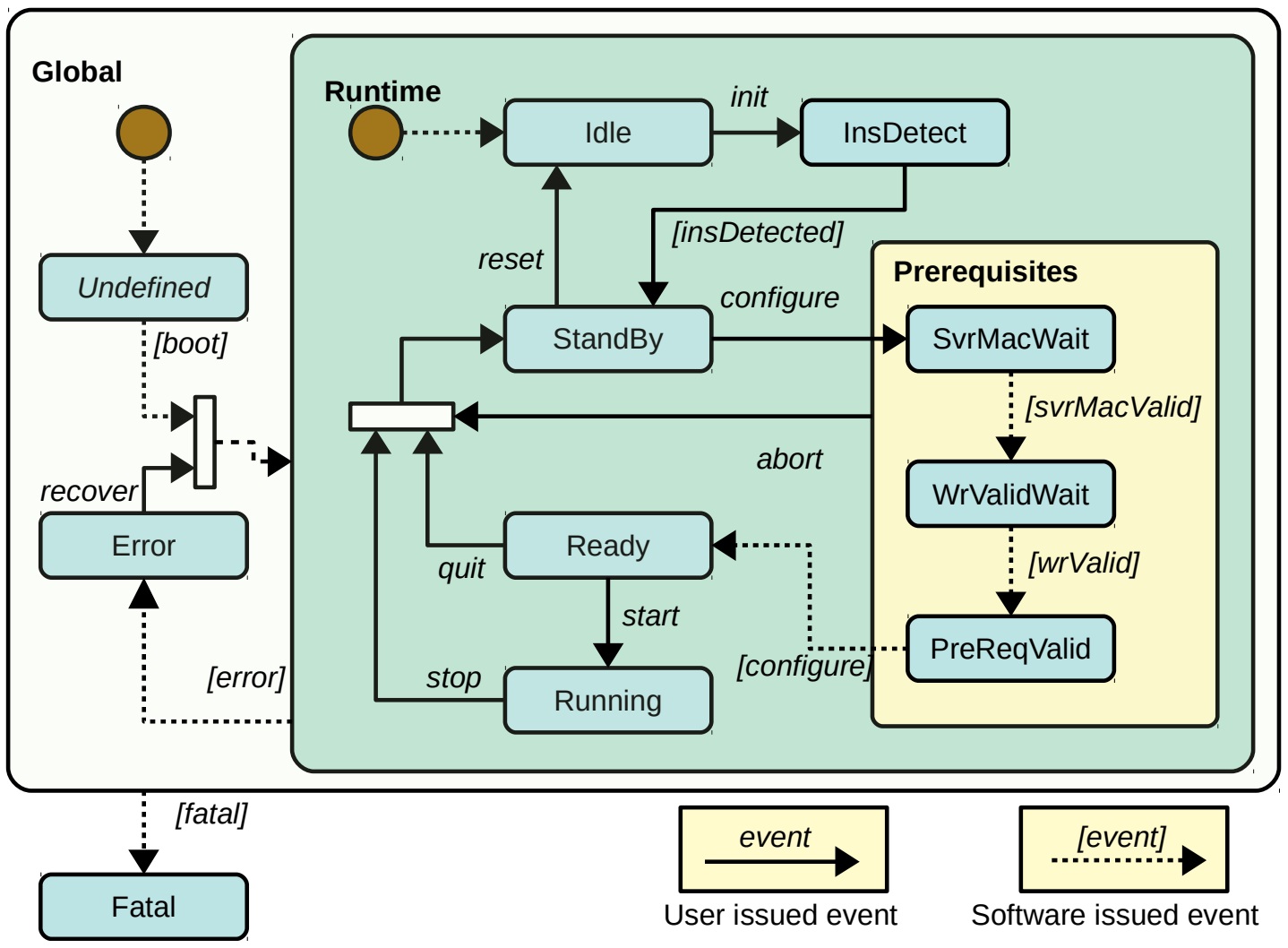

Figure 11: Block scheme of the software state machine. A description of the state machine is reported in the text.

The application layer of the CLB has been implemented as a state machine (shown in Fig. 11). The state machine receives events, either internally generated, or from remote, and executes the code belonging to the associated state transition. When started, the state machine is in the `Undefined` state and the internal `boot` event is automatically issued. When successful, it will cause the system to enter the `Idle` state. Any critical error during the `boot` event transition will cause the issue of the `fatal` event, and will cause the system transition to the `Fatal` state, which can only be left through a system power cycle or FPGA reconfiguration.

The events and states of the state machine are defined in `statemachine.json5`, and are used to generate the code required to run the state machine. Code can be executed on state entry, state exit, and state transition.

*9.3. Subsystems*

The application code is grouped into *subsystems*. A subsystem is a unit of code and data responsible for a specific aspect of CLB operation. It controls the hardware peripherals associated with this aspect, generally on state machine events. Additionally it provides a comprehensive interface to through

```
{ state      : "StandBy",
  entrystub : true,
  ordinal   : 2,
  transits  :
  { "reset"       : "Idle",
    "configure"   : "PreReq_Mac"
  }
}
```

Listing 7: Definition of the stand-by state, one of the ten states of the embedded software state-machine.

slow control commands and process variables. The different subsystems have been are described in Section 6.

For example, on the *start* event the state machine moves from `Ready` to `Running`. On this transition the System subsystem enables the data acquisition hardware to start data taking.

### *9.4. Typical operation*

The shore-located Detector Manager (DM), responsible for controlling the entire experiment, knows three high-level states: *Off*, *On* and *Run*. These states correspond to CLB states `Idle`, `StandBy` and `Running`, respectively ( Fig. 11). Generally, the detector is taking physics data and will be in the *Run* state. This means the DM will move each CLB to the `Running` state regardless of the state it is found in. A run typically lasts for several hours, after which the DM stops the acquisition by moving all CLBs to the `StandBy` state and then to `Ready` and `Running` again for the next run. In the last transition, the new run configuration [22] is applied to all the CLBs under control.

After start-up, if there are no errors, the CLB will be in the `Idle` state. Assuming the DM is not in *Off* state, it will send an `init` event to the CLB, causing the CLB to to transition and calling the associated init functions in each subsystems. This will initialize the hardware required for data acquisition. If successful, the CLB will enter the `StandBy` state. If the DM is in the *Run* state it will configure the CLB by first writing the configurable process variables, and then by issuing the `configure` event. This event will cause the CLB subsystems bound to this event to check the prerequisites running and configure required hardware. If successful, the CLB will be in the `Ready` state, which is only transitional from the DM perspective. The DM will immediately issue a `start` event, causing the CLB to enable the hardware state machine, initiating data taking and moving the CLB to the `Running` state.

When a run has ended, the DM will issue a `stop` event to all CLBs. When a new run is started, the CLB will be configured again, and will be brought into a running state as described previously. This sequence repeats during normal detector operation. The detector generally is only brought to *Off* during maintenance.

## 10. Conclusions

The embedded software of the KM3NeT acquisition electronics has been presented. In particular, the architecture of the embedded software has been described as well as the tools and methods for building the environment, and for generating the process variables. The embedded software has been successfully operated in DOMs and DU-Bases of almost 40 deployed DUs, with more than 20000 photomultipliers installed. It is also being operated in the tests of the last version of the CLB (v4) and the tests of new optical network architecture based on standard White Rabbit.

## Acknowledgements

The authors acknowledge the financial support of the funding agencies: Agence Nationale de la Recherche (contract ANR-15-CE31-0020), Centre National de la Recherche Scientifique (CNRS), Commission Européenne (FEDER fund and Marie Curie Program), LabEx UnivEarthS (ANR-10-LABX-0023 and ANR-18-IDEX-0001), Paris Île-de-France Region, France; Shota Rustaveli National Science Foundation of Georgia (SRNSFG, FR-22-13708), Georgia; The General Secretariat of Research and Innovation (GSRI), Greece Istituto Nazionale di Fisica Nucleare (INFN), Ministero dell'Università e della Ricerca (MIUR), PRIN 2017 program (Grant NAT-NET 2017W4HA7S) Italy; Ministry of Higher Education, Scientific Research and Innovation, Morocco, and the Arab Fund for Economic and Social Development, Kuwait; Nederlandse organisatie voor Wetenschappelijk Onderzoek (NWO), the Netherlands; The National Science Centre, Poland (2021/41/N/ST2/01177); The grant "AstroCeNT: Particle Astrophysics Science and Technology Centre", carried out within the International Research Agendas programme of the Foundation for Polish Science financed by the European Union under the European Regional Development Fund; National Authority for Scientific Research (ANCS), Romania; Grants PID2021-124591NB-C41, -C42, -C43 funded by MCIN/AEI/ 10.13039/501100011033 and, as appropriate, by “ERDF A way of making Europe”, by the “European Union” or by the “European Union NextGenerationEU/PRTR”, Programa de Planes Complementarios I+D+I (refs. ASFAE/2022/023, ASFAE/2022/014), Programa Prometeo (PROMETEO/2020/019) and GenT (refs. CIDEGENT/2018/034, /2019/043, /2020/049. /2021/23) of the Generalitat Valenciana, Junta de Andalucía (ref. SOMM17/6104/UGR, P18-FR-5057), EU: MSC program (ref. 101025085), Programa María Zambrano (Spanish Ministry of Universities, funded by the European Union, NextGenerationEU), Spain; The European Union's Horizon 2020 Research and Innovation Programme (ChETEC-INFRA - Project no. 101008324)

**Vincent van Beveren** is a PhD. candidate and embedded software engineer at National Institute for Subatomic Physics Nikhef. He received his BS in computer science and information technology in 2004 from the Hogeschool Utrecht. He has collaborated on multiple publications on various topics. His areas of expertise is in ultra-lower power embedded systems, digital signal processing and Internet of Things (IoT).

**Diego Real** is a PhD. in Physics and Research Engineer at Instituto de Física Corpuscular. He received his BS in Electronics in 1997 and his MS in Control and Electronics in 2000, both from the Polytechnic University of Valencia. He is the author of several publications on electronics. The Spanish Astronomy Academy awarded his PhD. with the Prize in the Instrumentation Category. His current research interests include acquisition and synchronisation systems for particle physics. He is, since 2013, the Electronics project leader of the KM3NeT telescope.